\documentstyle[preprint,aps]{revtex}

\catcode`\@=11 % This allows us to modify PLAIN macros.
\def\lsim{\mathrel{\mathpalette\@versim<}}
\def\gsim{\mathrel{\mathpalette\@versim>}}
\def\@versim#1#2{\vcenter{\offinterlineskip
        \ialign{$\m@th#1\hfil##\hfil$\crcr#2\crcr\sim\crcr } }}
\makeatletter
\def\@seccntformat#1{\csname the#1\endcsname.\hskip 1em}
\makeatother
\begin{document}

\newcommand{\z}{$Z^0$~}
\newcommand{\ep}{e$^+$e$^-$}
\begin{flushright}
{\footnotesize\renewcommand{\baselinestretch}{.75}
  SLAC--PUB--95--6969\\
September 1995\\
(T/E)
}
\end{flushright}

%%%%%%%%%%%%%%%%%%%%%%%%%%%%%%%%%%%%%%%%%%%%%
%%%%%%%%%%
%\input epsf
%------------------------------------------------------
% Title page
%------------------------------------------------------
%%%%%%%%%%%%%%%%%%%%%%%%%%%%%%%%%%%%%%%%%%%%%
%%%%%%%%%%%%%%%%%%%%%%%%%%%%%%%%%%
% 	Title & abstract
%%%%%%%%%%%%%%%%%%%%%%%%%%%%%%%%%%%%%%%%%%%%%
%%%%%%%%%%%%%%%%%%%%%%%%%%%%%%%%%%
\thispagestyle{empty}

\begin{center}

\centerline{\bf First Measurement of the T-odd Correlation between the}
\centerline{\bf \z Spin and the Three-jet Plane Orientation}
\centerline{\bf in Polarized \z Decays to Three Jets$^*$}

\vspace {1.0cm}

 {\bf The SLD Collaboration$^{**}$}\\
Stanford Linear Accelerator Center \\
Stanford University, Stanford, CA~94309

\vspace{1.0cm}

\end{center}

\normalsize

\begin{abstract}
 We present the first measurement of the correlation between the
 $Z^0$ spin and the three-jet plane orientation in
 polarized $Z^0$ decays into three jets in the SLD experiment
 at SLAC utilizing a longitudinally polarized electron beam.
 The CP-even and T-odd triple product
 $\vec{S_Z}\cdot(\vec{k_1}\times \vec{k_2})$
 formed from the two fastest jet momenta, $\vec{k_1}$  and $\vec{k_2}$,
 and the $Z^0$ polarization vector $\vec{S_Z}$, is sensitive to
 physics beyond the Standard Model.
 We measure the expectation value of this quantity to be consistent
 with zero and set 95\% C.L. limits of $-0.022 < \beta < 0.039$ on
 the correlation between the $Z^0$-spin and
 the three-jet plane \hbox{orientation}.

\vspace{0.5cm}

\centerline{Submitted to Physical Review Letters}

\vbox{
\uchyph=200%%Controls hyphenation%%%%%%%%%%%%
\brokenpenalty=200
%%Specifies penalty for pagebreak after discretionary item
\pretolerance=10000
%%Specifies tolerable badness for line breaks w/o hyphens
\tolerance=2000
%%Specifies tolerable badness for line breaks w hyphenation
\nobreak%%forces line not to break
\penalty 5000%%Neg encourages line break 10,000+ prevents break
\hyphenpenalty=5000%%Penalty for hyphens increase to discourage break
\exhyphenpenalty=5000
%Penalty for explicit hyphen Increase to discourage break
\footnotesize\renewcommand{\baselinestretch}{1}\noindent
$^*$This work was supported by Department of Energy
  contracts:
  DE-FG02-91ER40676 (BU),
  DE-FG03-92ER40701 (CIT),
  DE-FG03-91ER40618 (UCSB),
  DE-FG03-92ER40689 (UCSC),
  DE-FG03-93ER40788 (CSU),
  DE-FG02-91ER40672 (Colorado),
  DE-FG02-91ER40677 (Illinois),
  DE-AC03-76SF00098 (LBL),
  DE-FG02-92ER40715 (Massachusetts),
  DE-AC02-76ER03069 (MIT),
  DE-FG06-85ER40224 (Oregon),
  DE-AC03-76SF00515 (SLAC),
  DE-FG05-91ER40627 (Tennessee),
  DE-AC02-76ER00881 (Wisconsin),
  DE-FG02-92ER40704 (Yale);
  National Science Foundation grants:
  PHY-91-13428 (UCSC),
  PHY-89-21320 (Columbia),
  PHY-92-04239 (Cincinnati),
  PHY-88-17930 (Rutgers),
  PHY-88-19316 (Vanderbilt),
  PHY-92-03212 (Washington);
  the UK Science and Engineering Research Council
  (Brunel and RAL);
  the Istituto Nazionale di Fisica Nucleare of Italy
  (Bologna, Ferrara, Frascati, Pisa, Padova, Perugia);
  and the Japan-US Cooperative Research Project on High Energy Physics
  (Nagoya, Tohoku).}
\newpage

\normalsize

\end{abstract}

Polarization is an essential tool in
investigations of fundamental symmetries in particle  physics.
Parity violation was first discovered in $\beta$ decays from polarized
$^{60}$Co  \cite{Wu}, and
T, CP  and CPT violations were searched for using polarized
neutrons
\cite{Henley} and polarized positronium \cite{Arbic}. The recent
development of
high-polarization electron sources based on strained-lattice GaAs
photocathodes
\cite{Maruyama}, in conjunction with the high luminosity achieved
at the SLAC Linear
Collider (SLC), has allowed production of highly polarized $Z^0$ bosons
by e$^+$e$^-$ annihilation, enabling
investigations of symmetries at the $Z^0$ resonance.

The $Z^0$ bosons produced using
longitudinally polarized electrons have polarization
along the beam direction $A_Z = (P_{e^-}-A_e)/(1-P_{e^-}\cdot A_e)$,
where  $P_{e^-}$ is
the electron-beam polarization, defined to be negative (positive) for a
 left-(right-) handed
beam, and $A_e = 2v_ea_e/(v_e^2+a_e^2)$ with $v_e$ and $a_e$ the
electroweak vector and axial
vector coupling parameters of the electron, respectively. Since 1993
the SLC has run with a strained-lattice GaAs electron source;
 an electron-beam polarization at the \ep~interaction point
of approximately 0.77 in magnitude was achieved
in the 1994--95 run, yielding $A_Z$ = $-0.82$ ($+0.70$) for $P_{e^-}$ =
 $-0.77$ ($+0.77$) respectively, assuming $\sin^2\theta_w$=0.2319 $\pm$
 0.0005 \cite{PDG}. In order to reduce systematic effects,
the electron spin direction was randomly reversed  pulse-by-pulse, thus
achieving higher sensitivities to polarization-dependent asymmetries.
For polarized $Z^0$ decays to
three hadronic jets one can define the triple product:

\begin{equation}
 \vec{S_Z}\cdot(\vec{k_1}\times \vec{k_2}),
\label{triple}
\end{equation}

\noindent
which correlates the \z boson polarization vector $\vec{S_Z}$
with the normal to the three-jet plane defined by
$\vec{k_1}$ and $\vec{k_2}$, the momenta of the highest- and the
second-highest energy jets, respectively.
Here we report the first experimental study of this quantity.

The triple product (\ref{triple}) is even under C and P
 reversals, and odd
under T$_{\rm N}$, where T${\rm _N}$ reverses momenta and spin vectors without
exchanging initial
and final states. Since T$_{\rm N}$ is not a true time-reversal operation, a
non-zero value does
not signal CPT violation and is possible in a theory that
respects CPT
invariance \cite{Rujula71}. Similar observables were first proposed
for direct
experimental observation of the non-Abelian character of QCD in $e^+e^-
 \rightarrow \Upsilon \rightarrow ggg$ \cite{Rujula78},
and in $e^+e^- \rightarrow q{\bar q}g$
\cite{Fabricius} where a sizable signal is expected
at c.m. energies $\sqrt{s}$
below 40 GeV; no experimental measurements have been performed
since a  longitudinally polarized electron beam is required. A similar
triple product was studied theoretically in neutrino scattering
 \cite{Hagiwara81} and
lepton-nucleon scattering \cite{Hagiwara83}. More recently other
observables have  been proposed for high-energy jet physics to
explore
CP or T violation
\cite{Donoghue}.

The differential cross section for $e^+e^- \rightarrow q{\bar q}g$ for a
longitudinally polarized
electron beam and massless quarks may be written \cite{Fabricius}
\cite{Brandenburg}

\begin{equation}
{1 \over \sigma}
{{d\sigma} \over {d \cos\omega}} = {{9} \over {16}} [ (1-{1\over
 3}\cos^2\omega)+\beta
\, A_Z \, \cos\omega],
\label{diff}
\end{equation}

\noindent
where $\omega$ is the polar angle of the vector normal to the jet plane,
$\vec{k_1}\times \vec{k_2}$, w.r.t. the electron-beam direction.
With $|\beta A_Z|$ representing the magnitude
\cite{beta}, the second term is proportional to the T$_N$-odd
triple product
(\ref{triple}), and appears as a forward-backward asymmetry of the
jet-plane normal relative to the \z polarization axis.
The
sign and magnitude of this term are different for the two beam
helicities.
%the $\cos\omega$ distribution is examined separately for events
%produced by
%left- and right-handed beams, and the forward-backward asymmetry
%and the average
%value of $\cos\omega$ are evaluated in each case.

Recently Brandenburg, Dixon, and Shadmi have investigated Standard Model
T$_{\rm N}$-odd contributions of the form (\ref{triple})
at the \z resonance \cite{Brandenburg}. The
triple product vanishes identically at tree level \cite{Rujula71},
but  non-zero contributions arise
from higher-order processes such as those shown in Fig.~1:
(a) QCD rescattering of massive quarks \cite{Fabricius}, (b) QCD
triangle of massive quarks \cite{Hagiwara}, and (c) electroweak
rescattering via $W$ and
$Z$ exchange loops. Due to various cancellations these contributions
are found to be very small at the $Z^0$ resonance and yield values of
the correlation parameter
$ |\beta | \lsim 10^{-5}$~\cite{Brandenburg}.
Because of this background-free situation, measurement of the
cross section (\ref{diff}) is
sensitive to physics processes beyond the Standard Model that give
$\beta \neq$ 0.

 The measurement was performed with the SLC Large Detector (SLD) using
 approximately
50,000 \z decays into multi-hadrons collected in 1993 and 100,000 decays
collected
in 1994-95, for which the magnitude of the average electron-beam
polarization was 0.63 and 0.77 respectively. A general description of
 the SLD can
be found elsewhere~\cite{sld}. Charged particle tracking and momentum
analysis is provided by the central drift chamber (CDC)
%\cite{cdc}
and the CCD-based vertex detector
%\cite{vxd}
in a uniform axial magnetic field of 0.6~T. Particle energies are
measured in the liquid argon calorimeter (LAC) \cite{lac} and in the
warm iron calorimeter \cite{wic}.
Three triggers were used for hadronic events. The first required a
total LAC
electromagnetic energy greater than 12 GeV; the second required at
least two
well-separated tracks in the CDC; the third required at least 4 GeV
in the LAC and one track in the CDC.

In this analysis the hadronic event selection and three-jet
reconstruction were based on
the topology of energy depositions in the LAC, taking advantage of
its large solid-angle
coverage. The LAC is a lead liquid argon sampling calorimeter composed
of barrel and
endcap sections, covering the angular ranges $|\cos\theta|< 0.82$ and
$0.82 < |\cos\theta|
<~0.98$, respectively. It is segmented radially into projective towers
of constant solid angle with 192 azimuthal and 96 polar-angle
 segmentations. The longitudinal
segmentation comprises two electromagnetic sections with a combined
 thickness of 21 radiation
lengths, and two hadronic sections, giving a total thickness of 2.8
interaction lengths.

The calorimetric analysis must distinguish $Z^0$ events from
backgrounds;  in
addition it should remove any background hits coincident with $Z^0$
events. The
dominant source of beam-related backgrounds in the LAC was high-energy
muons produced
in the SLC that were characterized by small amounts of energy in a
large number of towers
parallel to the beam direction. An algorithm was used to identify this
 characteristic signal
and background hits were removed before the hadronic event selection
\cite{Alr}.

Although the LAC offers a uniform energy response over most
of its solid-angle
coverage, the response  is degraded around $|\cos\theta|
\approx$ 0.82, where the
barrel and endcap sections meet. In order to achieve a uniform
response over the
whole acceptance, the energy response of the towers was corrected.
The total detected energy was expressed as a  linear
combination of the tower energies weighted by energy-independent
constants

\begin{equation}
E_{detect} = \sum_{i} ( a_i \cdot E_{em}^i + b_i \cdot E_{had}^i),
\label{Edetect}
\end{equation}

\noindent
where $E_{em}^i$ and $E_{had}^i$ are the
recorded energies in the electromagnetic and
hadronic sections, and the sum was taken over all the polar-angle
segmentations \cite{phi};
$a_i$ and $b_i$ are correction factors determined by
minimizing the sum

\begin{equation}
\sum_{events}  {{(E_{detect}-E_{CM})^2} \over {\sigma^2}},
\label{Esum}
\end{equation}

\noindent
where $E_{CM}$ is the $e^+e^-$ collision energy
corrected for the detector acceptance and for
the undetectable energy carried by neutrinos, and $\sigma$ is the
measured LAC energy resolution for hadronic \z events as a function of
thrust axis \cite{thrust} polar angle $\theta^{thrust}$.
The sum was taken over recorded back-to-back two-jet events that form a
statistically-independent sample to the three-jet events used for this
study.

After applying the energy-response correction,
calorimeter towers were grouped into clusters
\cite{Youssef}. A cluster was selected if at least
two towers contributed, its energy $E_{cluster}$
was at least 100 MeV, and the energy
 correlation in the electromagnetic section
4$E_{em1}\cdot E_{em2}$/$(E_{em1}+E_{em2})^2 >$ 0.1, where
$E_{em1}$ and $E_{em2}$ are the detected energies in the front and back
electromagnetic sections, respectively \cite{Ecor}. Using the selected
clusters the total visible energy
$E_{vis}$, normalized energy imbalance $E_{imb}=
 |\sum{\vec{E}_{cluster}}|/E_{vis}$,
number of selected clusters $N_{cluster}$, and
$\cos\theta^{thrust}$ were calculated for each event, and
multi-hadron events were selected by requiring well-balanced events
containing large energy deposits and a large number of clusters, namely
$E_{vis} >$ 20 GeV,  $E_{imb} < $ 0.6,
and
$N_{cluster} \ge$ 9 for $|\cos\theta^{thrust}| <$ 0.8 and $N_{cluster}
\ge$ 12 for
$|\cos\theta^{thrust}| >$ 0.8. In total 50,144 events from the 1993 run
and 99,265  events from
the 1994--95 run were selected. The efficiency  for selecting hadronic
events was estimated to be $92\pm2\%$, with a background in the
selected sample of $0.4\pm 0.2\%$,
dominated by $Z^0 \rightarrow \tau^+ \tau^-$
and $Z^0$ $\rightarrow$ \ep events.

To measure the triple-product correlation for $e^+e^-
\rightarrow q{\bar q}g$, three-jet
events were selected and the three momentum vectors of the jets were
 reconstructed.
Although the parton momenta are not directly measurable, at $\sqrt{s}$
$\approx$ 91 GeV the partons usually
appear as well-collimated jets of hadrons. Jets were
reconstructed using the ``Durham'' jet algorithm \cite{Bethke}.
Planar three-jet events were
selected by requiring exactly three reconstructed jets to be found with a
jet-resolution parameter value of
$y_c$=0.005 \cite{ycut}, the sum of the angles
between the three jets to be greater than 358$^\circ$,
and that each jet contain at least two clusters.
A total of 44,683 events satisfied these
criteria and were subjected to further analysis.

Such jet algorithms accurately reconstruct
the parton directions
but measure the parton energies poorly \cite{threejet}.
Therefore, the jet energies were
calculated by using the measured jet directions and solving the
three-body kinematics assuming massless jets, and
were then used to label the jets such that $E_1 > E_2 > E_3$.
 The energy of jet 1, for example, is given by

\begin{equation}
E_1 = \sqrt{s} {{\sin\theta_{23}} \over
 {\sin\theta_{12}+\sin\theta_{23}+\sin\theta_{31}}},
\label{energy}
\end{equation}

\noindent
where $\theta_{kl}$ is the angle between jets $k$ and $l$.

Since the energy and angular resolutions of the jet reconstruction
procedure determine the
sensitivity of the present measurement, a Monte Carlo simulation of
hadronic $Z^0$
decays \cite{Sjostrand} combined with a simulation of the detector
response was used to
study the quality of the jet reconstruction. To account properly for
 beam-related backgrounds in the simulation, real calorimeter hits
taken by a random
trigger were overlaid on the simulated \z events.
These events were then subjected to
the same reconstruction, hadronic event selection, and three-jet analysis
 procedures as
the real data. For those events satisfying the three-jet criteria,
exactly three jets were
reconstructed at the parton level by applying the jet algorithm to
the parton momenta. The
three parton-level jets were associated with the three detector-level
jets by choosing the
combination that minimized the sum of the angular differences between the
 corresponding jets.  The directions and energies of
jets at the parton level were then compared with those for the
corresponding jets at the detector level. The average angles
between the parton-jet and detector-jet directions were 2.9$^\circ$,
4.0$^\circ$,
and 7.2$^\circ$, for the highest, medium, and lowest energy jets,
respectively. Although the detector-jet
energies were much degraded, the reconstructed energies agreed
 well with
the parton-jet energies; the r.m.s energy difference between parton  and
 detector jets was
2.8, 5.2, and 5.2 GeV for the highest, medium, and lowest
energy jet, respectively.

Since in this analysis
the vector normal to the jet plane is determined by the two highest
energy jets,
reconstruction of the correct jet-energy ordering is essential.
For a three-jet event whose jets are labeled according to
the parton-jet energy ordering, six detector-jet energy orderings
are possible. For the three cases where the energy ordering of
any two jets does not agree between parton and detector levels,
the direction of the
jet-plane normal vector is opposite between the parton level and
detector level and
$\cos\omega$  will be
measured with the wrong sign. The probability of this,
$P_{mis}(|\cos\omega|)$, was determined as a function of
$\cos\omega$ from Monte Carlo studies.
Although the three-jet rate was largest for
$y_c \approx$ 0.002,
the misassignment probability $P_{mis}$ was found to be smallest for $y_c
\approx$ 0.012. Combining
these two factors, the experimental sensitivity to the T$_{\rm N}$-odd
contribution was found highest for the $y_c$ value of 0.005
used in this analysis.
For this $y_c$ value $P_{mis}$ varied from 0.25 around cos$\omega$ = 0
to 0.21 as $|{\rm cos}\omega|\rightarrow 1$;
averaged over all $\cos\omega$,
$<P_{mis}(|\cos\omega|)>$ $\approx$ 0.22.

For each event the reconstructed jet vectors were used to determine
the vector normal to the jet plane and its polar angle $\omega$,
from which the measured distribution of cos$\omega$ was derived.
A bin-by-bin correction factor $\epsilon(|\cos\omega|)$,
for detector acceptance and initial-state radiation, was
determined from Monte Carlo simulations by taking the ratio of the
distribution at the parton level for an event sample generated without
initial-state radiation to the
distribution at the detector level for an event
sample generated with initial-state radiation and
subjected to the same reconstruction, selection, and anlysis as the data.
Figure 2 shows the corrected cos$\omega$ distribution
separately for left- and right-handed beam events in the 1994--95 data
sample. A T$_N$-odd contribution would appear as a forward-backward
asymmetry, of opposite sign between the left- and right-handed events;
no asymmetry is apparent. The distributions may be described by

\begin{equation}
{1 \over \sigma}
{{d\sigma} \over {d \cos\omega}} = {9 \over 16} [
 (1-{1\over
3}\cos^2\omega)+\beta \, A_Z \, (1-2\, P_{mis}(|\cos\omega|))\,
\cos\omega].
\label{diffc}
\end{equation}

We performed a maximum-likelihood fit of Eq. 6 simultaneously to the
cos$\omega$ distributions from the
1993 and 1994--1995 left- and right-handed event samples, with the relevant
values of $A_Z$, and allowing the parameter $\beta$ to vary.
We found
\begin{equation}
\beta = 0.008 \pm 0.015,
\label{beta}
\end{equation}

\noindent
where the error is statistical only \cite{bquark}. The result of this fit
is shown in Fig.~2; the $\chi^2$ is 26.0 for 20 data points.
The T$_N$-odd contribution is consistent with zero
within the statistical error and we calculate limits of
\begin{equation}
-0.022 < \beta < 0.039 \quad @ \quad 95\% \quad C.L.
\label{limit}
\end{equation}

A number of systematic checks was performed.
The analysis was performed on samples of Monte Carlo events in which no
T$_N$-odd effect was simulated, yielding $\beta$ consistent with zero
within $\pm$0.010, implying
that any analysis bias is less than $\pm$0.02 at 95\% C.L.
The dependence on the jet-resolution parameter was studied by varying
$y_c$ between 0.001 and 0.03, and in each case
the T$_N$-odd contribution was found to be
consistent with zero within the statistical error.
The analysis was also performed using the JADE jet algorithm
\cite{jade} and $y_c$=0.01. While
$P_{mis}$ was somewhat larger
than the value for
the Durham algorithm, 0.25 averaged over $|\cos\omega|$, the
experimental sensitivity was comparable
as a result of the larger three-jet rate \cite{sldalp}.
The T$_N$-odd contribution was found to be consistent with zero.
Finally, the analysis was performed using only charged tracks measured
in the CDC.
While the event sample was reduced to about 50\% of the calorimetric
sample as a result
of the smaller solid-angle coverage of the CDC, the charged tracks
provided an
independent basis for selecting and reconstructing three-jet events
\cite{sldalp}. The T$_N$-odd
contribution was again consistent with zero for the same range of $y_c$.

In conclusion, we have made the first measurement of the T$_N$-odd
 correlation in
polarized \z decays to three-jets. We find the correlation to be
consistent with zero and set
95\% C.L. limits on beyond-the-Standard-Model T$_N$-odd contributions to \z
decays to three-jets of $-0.022 < \beta < 0.039$.

We thank A. Brandenburg, L. Dixon, and Y. Shadmi for their efforts in
 calculating the
Standard Model prediction, and for enlightening discussions.
We thank the personnel of the SLAC accelerator department and the
technical
staffs of our collaborating institutions for their outstanding efforts
on our behalf.

\section*{$^{**}$List of Authors} %%%%%%% beginning of author list %%%%%%%%%%

\begin{center}
%
%   Institution number definitions:
%
  \def\iADEL{$^{(1)}$}
  \def\iBOL{$^{(2)}$}
  \def\iBU{$^{(3)}$}
  \def\iBRUN{$^{(4)}$}
  \def\iCIT{$^{(5)}$}
  \def\iUCSB{$^{(6)}$}
  \def\iUCSC{$^{(7)}$}
  \def\iCIN{$^{(8)}$}
  \def\iCSU{$^{(9)}$}
  \def\iCOLO{$^{(10)}$}
  \def\iCOL{$^{(11)}$}
  \def\iFER{$^{(12)}$}
  \def\iFRA{$^{(13)}$}
  \def\iILL{$^{(14)}$}
  \def\iLBL{$^{(15)}$}
  \def\iMIT{$^{(16)}$}
  \def\iMASS{$^{(17)}$}
  \def\iMISS{$^{(18)}$}
  \def\iNAG{$^{(19)}$}
  \def\iOREG{$^{(20)}$}
  \def\iPAD{$^{(21)}$}
  \def\iPERU{$^{(22)}$}
  \def\iPISA{$^{(23)}$}
  \def\iRUT{$^{(24)}$}
  \def\iRAL{$^{(25)}$}
  \def\iSOGANG{$^{(26)}$}
  \def\iSLAC{$^{(27)}$}
  \def\iTENN{$^{(28)}$}
  \def\iTOH{$^{(29)}$}
  \def\iVAND{$^{(30)}$}
  \def\iWASH{$^{(31)}$}
  \def\iWISC{$^{(32)}$}
  \def\iYALE{$^{(33)}$}
  \def\dead{$^{\dag}$}
  \def\andgen{$^{(a)}$}
  \def\andper{$^{(b)}$}
%
%  \author{                         % author and institution list
%  \baselineskip=.75\baselineskip   % shrink the interline spacing
%
\mbox{K. Abe                 \unskip,\iTOH}
\mbox{I. Abt                 \unskip,\iILL}
\mbox{C.J. Ahn               \unskip,\iSOGANG}
\mbox{T. Akagi               \unskip,\iSLAC}
\mbox{N.J. Allen             \unskip,\iBRUN}
\mbox{W.W. Ash               \unskip,\iSLAC$^\dagger$}
\mbox{D. Aston               \unskip,\iSLAC}
\mbox{K.G. Baird             \unskip,\iRUT}
\mbox{C. Baltay              \unskip,\iYALE}
\mbox{H.R. Band              \unskip,\iWISC
% ...
}
\it
  \vskip \baselineskip                   % \bigskip did not work
  \centerline{(The SLD Collaboration)}   % include collaboration name
  \vskip \baselineskip                   % \bigskip did not work
%
%  }   % end of author list
%
%  \address{                        % institution address list
  \baselineskip=.75\baselineskip   % shrink the interline spacing
  \iADEL
     Adelphi University,
     Garden City, New York 11530 \break
  \iBOL
     INFN Sezione di Bologna,
     I-40126 Bologna, Italy \break
  \iBU
     Boston University,
     Boston, Massachusetts 02215 \break
  \iBRUN
     Brunel University,
     Uxbridge, Middlesex UB8 3PH, United Kingdom \break
  \iCIT
     California Institute of Technology,
     Pasadena, California 91125 \break
  \iUCSB
     University of California at Santa Barbara,
     Santa Barbara, California 93106 \break
  \iUCSC
     University of California at Santa Cruz,
     Santa Cruz, California 95064 \break
  \iCIN
     University of Cincinnati,
     Cincinnati, Ohio 45221 \break
  \iCSU
     Colorado State University,
     Fort Collins, Colorado 80523 \break
  \iCOLO
     University of Colorado,
     Boulder, Colorado 80309 \break
  \iCOL
     Columbia University,
     New York, New York 10027 \break
  \iFER
     INFN Sezione di Ferrara and Universit\`a di Ferrara,
     I-44100 Ferrara, Italy \break
  \iFRA
     INFN  Lab. Nazionali di Frascati,
     I-00044 Frascati, Italy \break
  \iILL
     University of Illinois,
     Urbana, Illinois 61801 \break
  \iLBL
     Lawrence Berkeley Laboratory, University of California,
     Berkeley, California 94720 \break
  \iMIT
     Massachusetts Institute of Technology,
     Cambridge, Massachusetts 02139 \break
  \iMASS
     University of Massachusetts,
     Amherst, Massachusetts 01003 \break
  \iMISS
     University of Mississippi,
     University, Mississippi  38677 \break
  \iNAG
     Nagoya University,
     Chikusa-ku, Nagoya 464 Japan  \break
  \iOREG
     University of Oregon,
     Eugene, Oregon 97403 \break
  \iPAD
     INFN Sezione di Padova and Universit\`a di Padova,
     I-35100 Padova, Italy \break
  \iPERU
     INFN Sezione di Perugia and Universit\`a di Perugia,
     I-06100 Perugia, Italy \break
  \iPISA
     INFN Sezione di Pisa and Universit\`a di Pisa,
     I-56100 Pisa, Italy \break
  \iRUT
     Rutgers University,
     Piscataway, New Jersey 08855 \break
  \iRAL
     Rutherford Appleton Laboratory,
     Chilton, Didcot, Oxon OX11 0QX United Kingdom \break
  \iSOGANG
     Sogang University,
     Seoul, Korea \break
  \iSLAC
     Stanford Linear Accelerator Center, Stanford University,
     Stanford, California 94309 \break
  \iTENN
     University of Tennessee,
     Knoxville, Tennessee 37996 \break
  \iTOH
     Tohoku University,
     Sendai 980 Japan \break
  \iVAND
     Vanderbilt University,
     Nashville, Tennessee 37235 \break
  \iWASH
     University of Washington,
     Seattle, Washington 98195 \break
  \iWISC
     University of Wisconsin,
     Madison, Wisconsin 53706 \break
  \iYALE
     Yale University,
     New Haven, Connecticut 06511 \break
  \dead
     Deceased \break
  \andgen
     Also at the Universit\`a di Genova \break
  \andper
     Also at the Universit\`a di Perugia \break
\end{center}

\vfill
\eject

\newpage

%------------------------------------------------------
% Figures
%------------------------------------------------------
%======================
% Figure Captions
%======================
\section*{Figure captions }

\noindent
{\bf Figure 1}.
Representative Feynman diagrams of higher-order interactions with
non-vanishing
contributions to the triple product: (a) QCD rescattering
($m_q
\neq 0$ is required), (b) triangle diagram via quark annihilation
($m_q' \neq 0$ is required),
and (c) electroweak rescattering.

\vskip\baselineskip

\noindent
{\bf Figure 2}.
Polar-angle distribution of the jet-plane normal with respect to the
electron-beam direction
for the 1994--95 data sample with (a) left-handed and (b) right-handed
electron beam. The
solid curve is the best fit to the combined 1993 and 1994--95 data
samples.


\begin{thebibliography}{99}
\bibitem{beta}
The value of $\beta$ depends on the parameter that determines
the jet multiplicity of events, and is not a universal constant.
\bibitem{phi}
We assume that the energy response is azimuthally uniform and is
symmetric in polar angle.
\bibitem{Ecor}
The energy correlation is used to suppress clusters from background sources.
\end{thebibliography}
\end{document}